\documentclass{vldb}
\usepackage{graphicx}
\usepackage{balance}

\usepackage{todonotes}

\usepackage{lipsum}

\usepackage{braket}

\usepackage{paralist}

\usepackage{microtype}

\usepackage{mathptmx}

\usepackage[font=bf]{caption}
\usepackage{subcaption}

\usepackage[hidelinks,pdftitle={GraphVista: Interactive Exploration Of Large Graphs},pdfauthor={Marcus Paradies, Michael Rudolf, Wolfgang Lehner}]{hyperref}

\usepackage{xspace}
\newcommand{\abbrev}[1]{\textsc{\MakeLowercase{#1}}\xspace}
\newcommand{\propername}[1]{\textsc{#1}\xspace}

\usepackage{xcolor}

\newcommand{\BFS}{\abbrev{BFS}}
\newcommand{\DFS}{\abbrev{DFS}}

\newcommand{\GEM}{\abbrev{GEM}}
\newcommand{\SQL}{\abbrev{SQL}}
\newcommand{\LDBC}{\abbrev{LDBC}}
\newcommand{\JSON}{\abbrev{JSON}}
\newcommand{\HTTP}{\abbrev{HTTP}}
\newcommand{\QUBLE}{\abbrev{QUBLE}}
\newcommand{\SPARQL}{\abbrev{SPARQL}}

\newcommand{\GRAPHITE}{\propername{Graph\-ite}}
\newcommand{\GRAPHVISTA}{\propername{Graph\-Vista}}

\begin{document}

\title{GraphVista: Interactive Exploration Of Large Graphs}

\numberofauthors{1}
\author{
    \alignauthor Marcus Paradies\titlenote{Both authors are also affiliated with SAP SE.}\hspace{30pt}Michael
    Rudolf\titlenote{This}\hspace{30pt}Wolfgang Lehner\\
    \affaddr{Database Systems Group, TU Dresden, Germany}\\
    \email{m.paradies@sap.com}, \email{michael.rudolf01@sap.com}, \email{wolfgang.lehner@tu-dresden.de}
}

\maketitle

\begin{abstract}
The potential to gain business insights from graph-structured data through graph analytics is increasingly attracting companies from a variety of industries,
ranging from web companies to traditional enterprise businesses.
To analyze a graph, a user often executes isolated graph queries using a dedicated interface\textemdash a procedural graph programming interface or a
declarative graph query language.
The results are then returned and displayed using a specific visualization technique.
This follows the classical ad-hoc \emph{Query}$\rightarrow$\emph{Result} interaction paradigm and often requires multiple query iterations until an interesting
aspect in the graph data is identified.
This is caused on the one hand by the schema flexibility of graph data and on the other hand by the intricacies of declarative graph query languages.
To lower the burden for the user to explore an unknown graph without prior knowledge of a graph query language, visual graph exploration provides an
effective and intuitive query interface to navigate through the graph interactively.

We demonstrate \GRAPHVISTA, a graph visualization and exploration tool that can seamlessly combine ad-hoc querying and interactive graph exploration within the
same query session.
In our demonstration, conference attendees will see \GRAPHVISTA running against a large real-world graph data set.
They will start by identifying entry points of interest with the help of ad-hoc queries and will then discover the graph interactively through visual graph
exploration.
\end{abstract}

\section{Introduction}
\label{sec:introduction}

\noindent
Traditional data exploration follows a direct, ad-hoc interaction with the database by using a declarative query language, such as \SQL, through a
console-oriented interface.
Also, the current generation of data interfaces---forms, reporting tools, and query workflows---is built based on this
\emph{Query}$\rightarrow$\emph{Result} paradigm:
an application extracts all the necessary information to construct a query, sends the query request to the database engine, and finally presents the received
result to the user~\cite{Nandi2013a}.
During query execution, however, no interaction with the database engine is possible.
Being able to evaluate the quality of the query results early on whenever a partial result is encountered, improves the user experience due to the increased
interaction.

With the ever-growing prevalence of large volumes of graph-structured data in enterprise applications, there is also an increasing demand for analyzing and
understanding the data to gain new insights---that includes not only the mere structure of the graph but also the role of attributes on vertices and edges.
At this scale, visualizing graph-structured data as-is does not only exceed human information reception capacity but also the resolution of the majority of
computer displays.
To reduce the information overload and aid the user in navigating the data, some tools~\cite{Abello2006, Akoglu2012} apply graph clustering and
summarization techniques (top-down) while others~\cite{Chau2011} employ query-by-example and machine learning techniques (bottom-up).
However, these approaches focus more on visualization than on graph querying and interactive exploration.

Application developers and data scientists leverage imperative programming interfaces and declarative graph query languages---such as \GEM~\cite{Rudolf2013},
\SPARQL and Cypher\footnote{\url{http://neo4j.org}}---to explore the graph topology and vertex/edge attributes.
Although these database interfaces provide an intuitive graph abstraction and a rich set of functionality, for many users in real-life domains (e.g., life
sciences) they impose a shallow learning curve, so that formulating or programming a graph query requires a considerable cognitive effort.
Also, they follow the traditional database interaction paradigm and construct the final result ad-hoc and in a single operation---interactive graph exploration
is not possible.

Many applications benefit from the schema flexibility that graph data models offer.
Unfortunately, querying and exploring a graph with an unknown schema (e.g., vertex and edge types, available attributes) is error-prone and often requires
multiple iterations until a meaningful result is returned.
Whereas \emph{Why-Empty} queries~\cite{Vasilyeva2014} can help explain empty results, keyword searches~\cite{Wang2010}, and query transformations using
abbreviations, synonyms, and ontologies~\cite{Yang2014} can simplify schema-agnostic querying.
However, these approaches follow the traditional query paradigm and do not permit interactive graph exploration.
The querying system \QUBLE interleaves visual query formulation of subgraph queries with the actual query processing to lower the system response
time~\cite{Bhowmick2013, Hung2014}, but it does not allow interactive exploration or iterative query processing of large graphs.


To address these challenges, the user should be offered a different query paradigm that allows the seamless combination of ad-hoc querying and
interactive graph exploration within a single query session.
More specifically, we propose that a graph query intent should capture two aspects:
\begin{inparaenum}[(1)]
\item a general specification of the query for the system with entry points where the query processing pauses, and
\item an interactive graph exploration starting from those entry points performed by the user at run-time.
\end{inparaenum}
The first part of the exploration is then an ad-hoc query, but instead of returning the complete result set eventually, the query immediately pauses the
execution after a matching vertex/edge was found and presents it to the user.
The second phase of the interaction allows using each partial result of the ad-hoc query as an entry point for the subsequent interactive graph exploration.

In this demonstration, we propose \GRAPHVISTA as a flexible tool for the analysis and visual exploration of large graphs.
It does not require a deep understanding of graph query languages or knowledge about the schema of the graph.
\GRAPHVISTA offers an extended query paradigm that seamlessly combines ad-hoc querying with interactive graph exploration to enable investigating the graph
starting from so-called \emph{graph query breakpoints}.
In analogy to software debugging, a graph query breakpoint is a single match of the initial ad-hoc query and serves as the entry point for an interactive
analysis of the graph.
\GRAPHVISTA also allows collecting intermediate results in the form of \emph{graph bookmarks}, which can be accessed and further explored later on.
Once the user has finished exploring the graph at the specific result, the execution of the ad-hoc query continues and presents the next result to the user.
This interleaved execution of ad-hoc querying to retrieve the entry points and visual, interactive graph exploration is a powerful paradigm and works equally
well for small and large graphs.

\section{GraphVista}
\label{sec:system}

\noindent
In this section we outline the architecture of \GRAPHVISTA---a browser-based graph visualization and exploration tool for large graphs---and describe its main
functionality in detail.

\begin{figure}[t!]
\centering
\includegraphics[width=\linewidth]{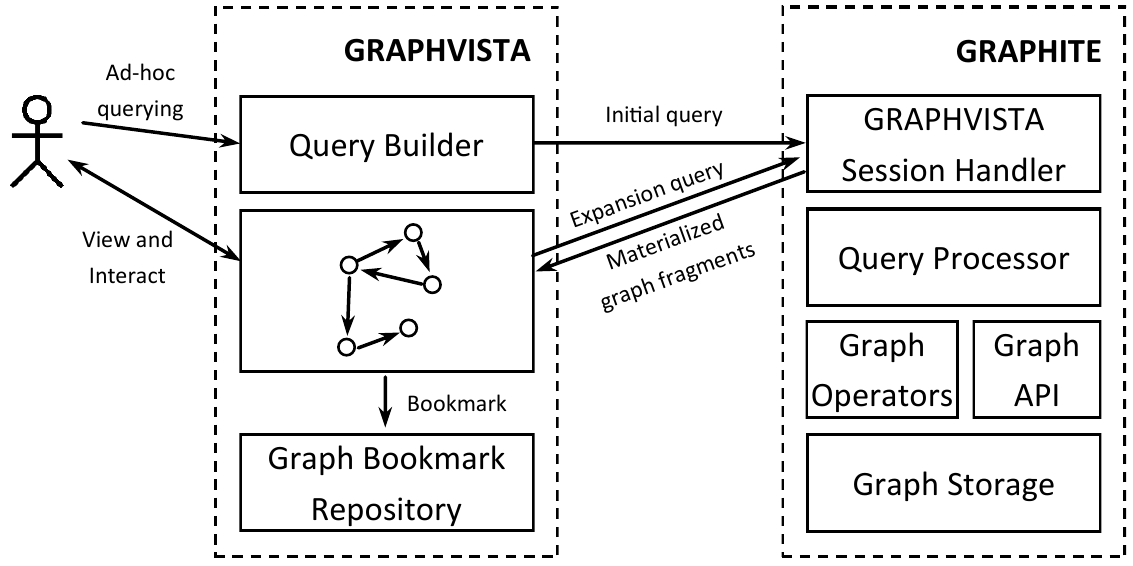}
\vspace{-15pt}
\caption{System architecture and integration of \GRAPHVISTA.}
\vspace{-10pt}
\label{fig:system_architecture}
\end{figure}

\subsection{System Overview}

\noindent
\GRAPHVISTA is a light-weight, visual query and exploration tool for inspecting and analyzing large graphs in an interactive manner and without the need to
formulate textual graph queries.
It aims at seamlessly blending ad-hoc querying with fine-granular iterative graph exploration manually performed by the user.
\GRAPHVISTA does not only allow querying the graph, it also allows storing intermediate results---so-called \emph{graph bookmarks}---in a repository for later
use.
Our goal is to allow the user analyzing large graphs at scale with the help of visual querying but without overloading the application by pulling in large
fractions of the data set.
We incorporate ad-hoc querying to skip those parts of the graph that are irrelevant for the analysis and to pause the query execution at user-defined
breakpoints allowing the user to interactively inspect the graph from there.

The overall architecture of \GRAPHVISTA is depicted on the left-hand side of Figure~\ref{fig:system_architecture}.
It consists of three major components: a \emph{query builder}, a \emph{graph visualization panel}, and a \emph{graph bookmark repository}.
The query builder provides graphical control elements for specifying the initial ad-hoc query and a set of \emph{graph query breakpoints}.
Just like a program's control flow steers a software debugging session, the ad-hoc query drives the graph analysis---we will therefore refer to it as
\emph{driver query}.
A graph query breakpoint is defined as a conjunctive predicate that is evaluated for each record that the driver query processes.
The interactive query panel contains graphical control elements to start, continue, or stop the query execution.
It also displays an excerpt from the entire graph data set, depending on the executed driver query and the performed interactive exploration steps.
The graph bookmark repository can be used to store the displayed parts of the entire graph for later use.
A graph bookmark consists of a concise description of the displayed vertices, edges, and accompanying attribute values.

\begin{figure}[t!]
\centering
\includegraphics[width=\linewidth]{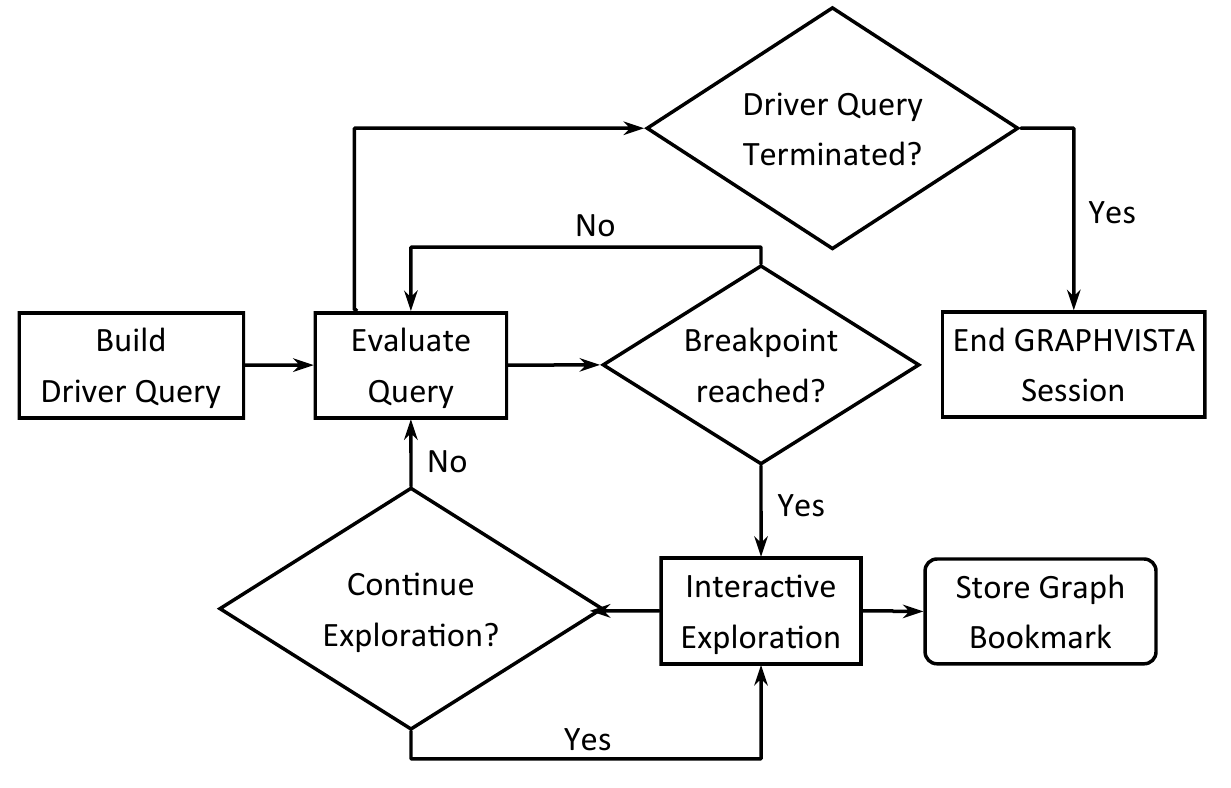}
\vspace{-10pt}
\caption{Interaction workflow of \GRAPHVISTA.}
\vspace{-10pt}
\label{fig:workflow}
\end{figure}

Figure~\ref{fig:workflow} illustrates the typical workflow with \GRAPHVISTA.
The user specifies the initial driver query and a set of graph query breakpoints using the query builder and starts the query execution.
Once a graph breakpoint is triggered, the driver query is paused and we switch from \emph{automated exploration} to \emph{interactive visual exploration}.
\GRAPHVISTA supports not only the display of subgraphs, but also the interactive exploration based on the already retrieved parts of the entire data graph.
The user can explore the result that triggered the breakpoint in a stepwise manner by fetching additional attribute values or by expanding the subgraph on the
incoming/outgoing edges.
The retrieved data can be saved as a graph bookmark for further exploration later on.
Once finished with the interactive exploration, the user can either continue the driver query---until the execution reaches the next graph breakpoint---or stop
its execution.

\GRAPHVISTA runs on top of \GRAPHITE~\cite{Paradies15}, a fast in-memory graph processing engine for interactive and analytic graph queries.
Its architecture is sketched on the right-hand side of Figure~\ref{fig:system_architecture}.
\GRAPHITE combines a columnar storage layer with native graph operators, light-weight secondary graph index structures, and a low-level graph programming
interface.
We store a graph in \GRAPHITE in two column groups---one for vertices and one for edges---and map each distinct attribute to a separate column.
Additionally, the graph topology can be either stored as an edge list represented by two columns or more efficiently in a high-performance adjacency list
structure with logical pointers to the corresponding rows in the vertex and edge column groups.

\begin{figure*}[t!]
\centering
\begin{subfigure}{0.48\textwidth}
    \includegraphics[width=\textwidth]{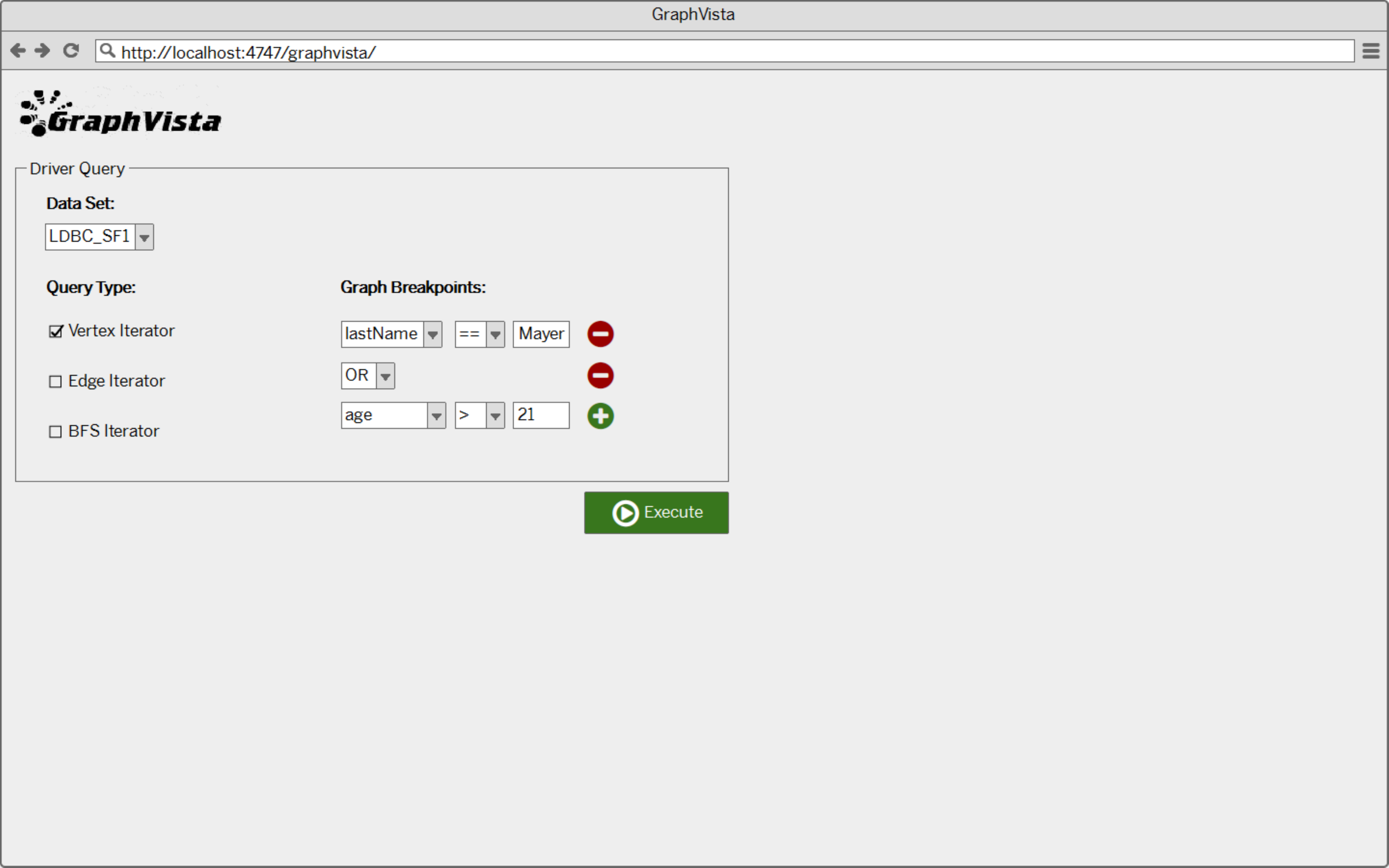}
    \caption{Driver query panel.}
    \label{fig:graphvista_driver_interface}
\end{subfigure}
\hfill
\begin{subfigure}{0.48\textwidth}
    \includegraphics[width=\textwidth]{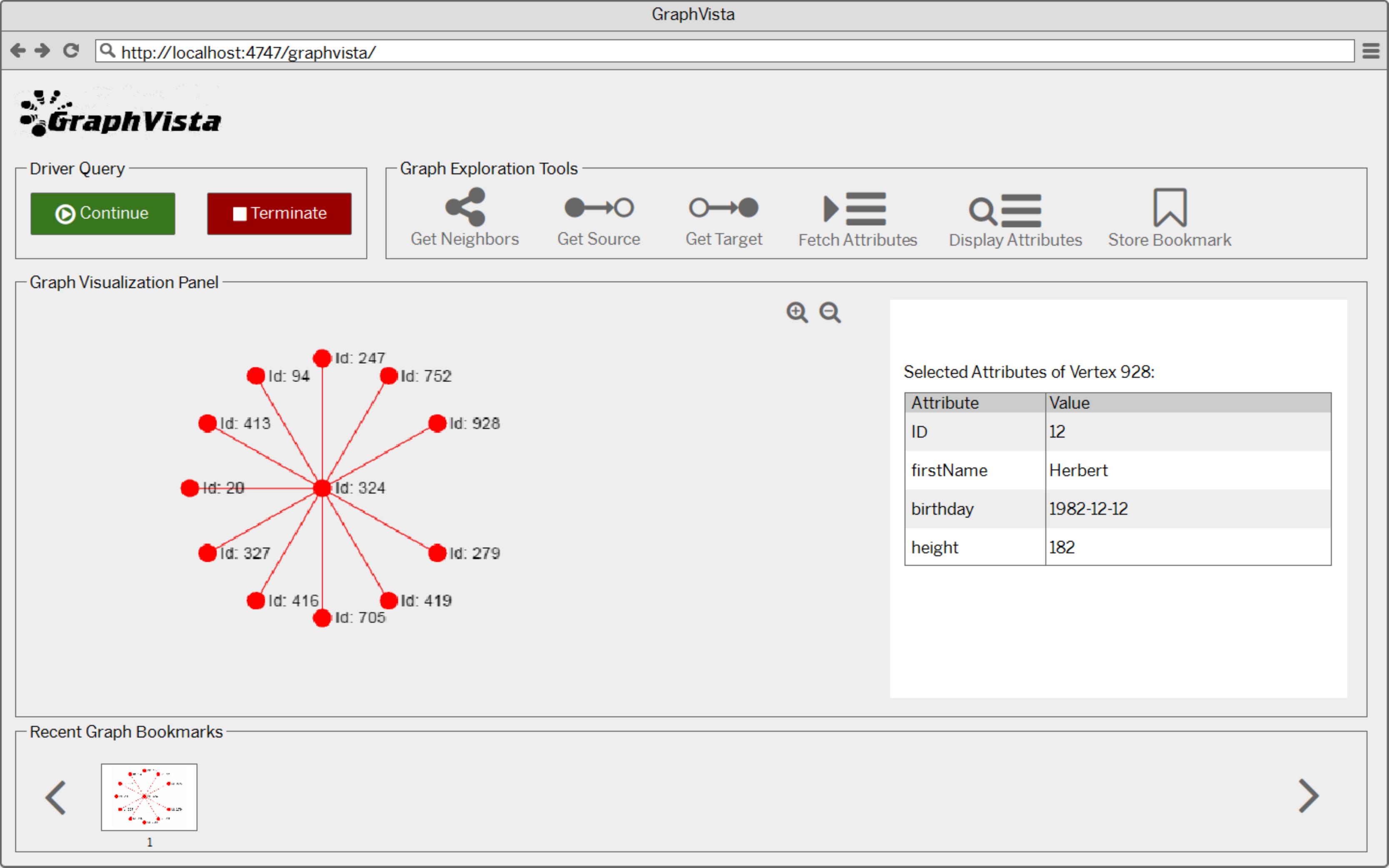}
    \caption{Interactive query panel.}
    \label{fig:graphvista_interactive_interface}
\end{subfigure}
\caption{\GRAPHVISTA graphical user interface.}
\label{fig:graphvista_user_interface}
\end{figure*}

\subsection{Automated Graph Exploration}

\noindent
A graph exploration in \GRAPHVISTA is composed of a driver query with an accompanying set of graph breakpoints and a series of graph exploration operations
performed by the user.
To formulate a driver query, we use a graphical query builder that is able to construct queries of two different classes:
\begin{inparaenum}[(1)]
  \item iterating over all vertices/edges---possibly restricted by predicates on attribute values---and
  \item iterating over vertices and edges in \BFS/\DFS order.
\end{inparaenum}
An active graph breakpoint causes the driver query to pause the execution in \GRAPHITE and to return the vertex/edge that triggered the breakpoint to
\GRAPHVISTA.
For example, the breakpoint condition $(\text{age}>21)$ is evaluated for each vertex/edge and returns a matching vertex/edge to
\GRAPHVISTA for display on the graph visualization panel.

Initially, we submit the driver query and the graph breakpoint conditions as additional filter predicates to the \GRAPHITE query processor.
As a result, we receive a unique session identifier---\GRAPHITE can handle multiple \GRAPHVISTA sessions simultaneously.
Conceptually, a driver query is executed in the same way as an ordinary query, but the predicate evaluation provides a callback mechanism that returns the
corresponding vertex/edge for each record satisfying the breakpoint condition.
If the user continues the driver query, \GRAPHITE fetches the intermediate result of the driver query and continues the execution until the next graph
breakpoint is reached.
The driver query terminates when either all vertices/edges have been visited or the \BFS/\DFS traversal terminates.

\subsection{Interactive Graph Exploration}

\noindent
One of the major advantages of \GRAPHVISTA is the ability to interleave the execution of the driver query with an interactive graph exploration initiated by the
user.
This allows skipping irrelevant parts of the graph during a traversal or vertex/edge iteration and only return vertices/edges of interest to the user, from
which the user can start the interactive graph exploration.
\GRAPHVISTA switches the execution mode from the driver query to interactive graph exploration whenever the graph breakpoint condition matches and displays the
respective vertex/edge on the graph visualization panel.
We support four kinds of operations to explore the graph in \GRAPHVISTA through graphical control elements in the graph visualization panel: 
\begin{inparaenum}[(1)]
  \item the retrieval of incoming/outgoing edges for a given vertex,
  \item the retrieval of preceding/succeeding vertices for a given edge/vertex,
  \item the retrieval of specific vertex/edge attributes, and
  \item the display of specific vertex/edge attributes.
\end{inparaenum}

For each exploration operation performed by the user, \GRAPHVISTA constructs a query request and dispatches it to the query processor of \GRAPHITE.
Each session in \GRAPHVISTA is identified by a unique session identifier that is used to tie the paused driver query to all subsequent interactive exploration
operations.
We use remote procedure calls to translate user exploration operations in \GRAPHVISTA to the corresponding operations in \GRAPHITE.
To return the result, \GRAPHITE constructs a response message containing the query result in the \JSON serialization format and transfers the response over an
\HTTP connection back to \GRAPHVISTA.
We then construct a new subgraph by merging the already displayed graph with the newly retrieved graph data and display it on the graph visualization panel.

Interactive graph exploration in \GRAPHVISTA is an iterative process and for each intermediate result presented by the driver query, the user can execute
different exploration actions to focus on the relevant parts of a potentially large graph on a single graph visualization panel.
The driver query either terminates once all vertices/edges have been presented to the user or the user terminates the driver query manually to start a different
driver query.
The user can decide to proceed with the driver query execution to fetch the next matching vertex/edge by issuing a continuation request to \GRAPHITE.
To handle skewed neighborhood size distributions efficiently and to not overload the graph visualization with too many vertices/edges, we allow restricting the
neighborhood expansion by the following three criteria:
\begin{inparaenum}[(1)]
  \item the direction of the edge,
  \item an edge attribute filter, and
  \item a vertex attribute filter.
\end{inparaenum}
To aid the user in applying these filters, we display an estimated result size before fetching the neighbor vertices.

\subsection{Graph Bookmarks}

\noindent
\GRAPHVISTA allows storing intermediate results during the interactive exploration of the fetched and displayed excerpts of the graph as bookmarks in a
repository.
A graph bookmark consists of the timestamp of creation, the materialized graph data (i.e.\ the explored topology and fetched vertex/edge attributes), and an
optional textual description provided by the user.
Conceptually, graph bookmarks are inspired by the Manylines project\footnote{\url{http://tools.medialab.sciences-po.fr/manylines}} but we extend their idea to
also operate on large graphs in conjunction with a graph processing system.
If the underlying graph changes between storing and retrieving the graph bookmark, fetching attributes from a deleted vertex/edge or retrieving
the neighbors of a deleted vertex results in a warning message that is displayed to the user.

\noindent\textbf{Storage:} We associate graph bookmarks with the session they were taken in and store them in \GRAPHVISTA in \JSON format containing
all retrieved information (i.e.\ metadata and data), so that they can be loaded quickly and without having to issue a query to the underlying graph engine for
collecting the data again.
Since graph bookmarks only contain the results of the interactive exploration and not a snapshot of the complete data set, storing them fully materialized
at the client-side is a feasible approach.

\noindent\textbf{Retrieval:} A graph bookmark can be retrieved in two ways:
\begin{inparaenum}[(1)]
  \item during the interactive exploration by selecting the graph bookmark of interest from the panel at the bottom, or
  \item in a separate session that can be started from the initial dialog of \GRAPHVISTA.
\end{inparaenum}
Once the user selects the graph bookmark, the contained vertices and edges are displayed on the graph visualization panel and the user can continue the
graph exploration from there.
Retrieving a bookmark and being able to use it as input or as initial starting point for another exploration is a powerful mechanism, because it allows
sifting the graph in multiple iterations instead of limiting the graph exploration to a single query session.

\subsection{Graphical User Interface}

\noindent
We developed \GRAPHVISTA with three major design goals in mind:
\begin{inparaenum}[(1)]
  \item it should be intuitive to use without requiring expert knowledge of a graph query language,
  \item it should be able to process large graphs while being able to explore the graph interactively, and
  \item it should be possible to use \GRAPHVISTA on gesture-based mobile devices.
\end{inparaenum}

We graphically separate the construction of the driver query and the interactive graph exploration into two window panels---a driver query panel
(cf.~Figure~\ref{fig:graphvista_driver_interface}) and an interactive query panel (cf.~Figure~\ref{fig:graphvista_interactive_interface}).
The driver query panel is composed of graphical control elements to select the data set to consider, the type of the driver query, and to specify the graph
breakpoints.
The interactive query panel is composed of graphical control elements to steer the driver query execution, a set of graph exploration operations visualized
through icons, the graph visualization panel, and a list of chronologically ordered thumbnails representing the stored graph bookmarks.
Next to the graph visualization panel, \GRAPHVISTA allows displaying previously fetched attributes of a selected vertex/edge in a tabular fashion.

\section{Demonstration}
\label{sec:demonstration}

\noindent
During the demonstration session, we will be running \GRAPHVISTA on top of \GRAPHITE and showcase an end-to-end solution for fast and intuitive interaction with
and exploration of large graph data sets.
The graphical user interface of \GRAPHVISTA is depicted in Figure~\ref{fig:graphvista_user_interface}.
We will use a social network data set generated from the \LDBC social network benchmark\footnote{\url{http://ldbcouncil.org/}} consisting of about 3.7~million
vertices and 21~million edges and load it into \GRAPHITE.
The data will be stored in two relational tables, one for vertices and one for edges, and an additional adjacency list data structure with logical pointers
to the corresponding vertex and edge entries in the tables.
The \LDBC data set models a social media website and knowledge exchange platform, where users can register, compose posts in forums, comment on posts by other
users, and become friends with other users.
The schema of the \LDBC data set distinguishes 8 different vertex types, 13 edge types, and a variety of descriptive attributes on vertices and edges.

\noindent\textbf{Scenario:}
If only a fraction of the graph schema is known to the user upfront and the query intent is not yet fully determined at the time of query formulation,
interactive graph exploration is a powerful approach.
In the demonstration, conference attendees will be able to specify driver queries while only being aware of parts of the graph schema, such as the available
vertex types.
For example, the conference attendee can specify a driver query that iterates over all female persons who live in the United States, are older than 21 years,
and joined the social network after 2009.
\GRAPHVISTA issues the query to \GRAPHITE and presents the first result---the first person that matches the selection criteria---on the graph visualization
panel.
Then the conference attendee can either continue the driver query to proceed to the next result or start the interactive graph exploration.
For example, the user might be interested in the 1-hop neighborhood and can restrict the neighborhood query further to only traverse edges of type
\emph{friendOf} to connected persons that are female.
Based on the retrieved neighborhood, the user can fetch additional attributes, such as \emph{firstname} and \emph{lastname}, of selected vertices.
At any time, the user can create graph bookmarks to view and extend the graph exploration at a later point in time.
In case the specified driver query is not sufficiently selective, the user can terminate its execution and start over again with a different driver query
configuration.
\GRAPHVISTA does not require the conference attendee to know the entire graph schema, but provides intuitive control elements and selection dialogs guiding the
user through the graph exploration process.
Furthermore, by fetching vertex neighborhood and vertex/edge attributes on demand, \GRAPHVISTA can operate on large graphs without having to load or display the
entire data set.

\balance

\bibliographystyle{abbrv}
\bibliography{bibliography}

\end{document}